\documentclass[8.5pt,twoside,twocolumn]{article}
\usepackage[super,sort&compress,comma]{natbib} 
\usepackage{times,mathptmx}
\usepackage{sectsty}
\usepackage{balance} 

\usepackage{graphicx} 
\usepackage{lastpage}

\usepackage{amsmath,amssymb,amsfonts,mathrsfs}
\usepackage{epic,eepic}
\usepackage{graphics}
\usepackage{epsfig}

\usepackage{subfigure}

\newcommand{\be}{\begin{equation}}
\newcommand{\ee}{\end{equation}}
\usepackage{amsmath}
\usepackage{bm}
\usepackage{amssymb}
\usepackage{natbib}
\usepackage{color}

\usepackage[format=plain,justification=raggedright,singlelinecheck=false,font=small,labelfont=bf,labelsep=space]{caption} 
\usepackage{fancyhdr}
\begin{document}
\thispagestyle{plain}
\fancypagestyle{plain}{
\renewcommand{\headrulewidth}{1pt}}
\renewcommand{\thefootnote}{\fnsymbol{footnote}}
\renewcommand\footnoterule{\vspace*{1pt}%
\hrule width 3.4in height 0.4pt \vspace*{5pt}} 
\setcounter{secnumdepth}{5}
\makeatletter 
\def\subsubsection{\@startsection{subsubsection}{3}{10pt}{-1.25ex plus -1ex minus -.1ex}{0ex plus 0ex}{\normalsize\bf}} 
\def\paragraph{\@startsection{paragraph}{4}{10pt}{-1.25ex plus -1ex minus -.1ex}{0ex plus 0ex}{\normalsize\textit}} 
\renewcommand\@biblabel[1]{#1}            
\renewcommand\@makefntext[1]%
{\noindent\makebox[0pt][r]{\@thefnmark\,}#1}
\makeatother 
\renewcommand{\figurename}{\small{Fig.}~}
\sectionfont{\large}
\subsectionfont{\normalsize} 

\fancyfoot{}
\fancyfoot[RO]{\footnotesize{\sffamily{1--\pageref{LastPage} ~\textbar  \hspace{2pt}\thepage}}}
\fancyfoot[LE]{\footnotesize{\sffamily{\thepage~\textbar\hspace{3.45cm} 1--\pageref{LastPage}}}}
\fancyhead{}
\renewcommand{\headrulewidth}{1pt} 
\renewcommand{\footrulewidth}{1pt}
\setlength{\arrayrulewidth}{1pt}
\setlength{\columnsep}{6.5mm}
\setlength\bibsep{1pt}

\twocolumn[
  \begin{@twocolumnfalse}
\noindent\LARGE{\textbf{Regularization of the slip length divergence in water nanoflows by inhomogeneities at the Angstrom scale}}
\vspace{0.6cm}

\noindent\large{\textbf{Marcello Sega,$^{\ast}$\textit{$^{a,b}$} Mauro Sbragaglia,\textit{$^{b}$} Luca Biferale,\textit{$^{b}$} and
Sauro Succi\textit{$^{c}$}}}\vspace{0.5cm}


\vspace{0.6cm}

\noindent \normalsize{
We performed non-equilibrium Molecular Dynamics simulations
of water flow in nano-channels with the aim of discriminating {\it
static} from {\it dynamic} contributions of the solid surface to
the slip length of the molecular flow.  We show that the regularization
of the slip length divergence at high shear rates, formerly attributed
to the wall dynamics, is controlled instead by its static properties.
Surprisingly, we find that atomic displacements at the Angstrom
scale are sufficient to remove the divergence of the
slip length and realize the no-slip condition. Since surface thermal
fluctuations at room temperature are enough to generate these
displacements, we argue that the no-slip condition for water can be achieved
also for ideal surfaces, which do not present any surface roughness.}

\vspace{0.5cm}
 \end{@twocolumnfalse}
  ]

\section{The debate about slip length divergence}


\footnotetext{\textit{$^{a}$~{Institut f\"ur Computergest\"utzte Biologische Chemie, W\"ahringer Strasse 17, 1090 Vienna, Austria}, e-mail: marcello.sega@univie.ac.at}}
\footnotetext{\textit{$^{b}$~{Department of Physics and INFN, University of Rome ``Tor Vergata'', Via della Ricerca Scientifica 1, I-00133 Rome, Italy} }}
\footnotetext{\textit{$^{c}$~{Istituto per le Applicazioni del Calcolo CNR, Viale del Policlinico 137, I-00161 Rome, Italy} }}


In the pioneering paper\cite{thompson97}, it was shown that the
Navier slip boundary condition $u_s= \ell_s \dot{\gamma}$ can be
regarded as the low-shear limit of a universal, nonlinear relation
between slip velocity $u_s$, the slip length $\ell_s$, and the local
shear rate $\dot{\gamma}$, that presents a divergence of the slip length upon approaching a critical value of the shear rate.
The onset of divergence is observed at
$\dot{\gamma}\simeq 10^{10}$~s$^{-1}$, both for argon\cite{thompson97}
and for water\cite{this}, a value which is accessible so far only to
numerical simulations, although the interest in the regime of
ultrahigh shear rates is rapidly growing: commercial, industrial-grade
fixed-geometry fluid processors like the microfluidizer (Microfluidics
Corp, MA, USA) operate at shear rates exceeding $10^{7}$~s$^{-1}$
to achieve uniform particle size reduction in emulsions and
dispersions, to create nanoencapsulations or to produce cell rupture.
Rheometers such as the piezoaxial vibrator\cite{crassous05} and the
torsion resonator\cite{fritz03} are also employed to measure the viscosity
of fluids at rates up to $10^{7}$~s$^{-1}$, e.g., for the characterization
of ink jet fluid rheology\cite{moseler00,vadillo10}. Laser-induced shock-waves
in confined water\cite{pezeril11}, with speed up to Mach 6~in a
5~$\mu$m thick water slab yield even larger numbers, which are of
the order of $10^{9}$~s$^{-1}$. A more detailed knowledge of the
properties of fluids at ultrahigh shear rates has therefore become
a compelling task. 
\begin{figure}
\makebox[0pt]{\raisebox{-0.4in}[0pt][0pt]{\raggedleft(a)}}
\includegraphics[bb=185 0 430 842,clip,angle=270,width=0.95\columnwidth]{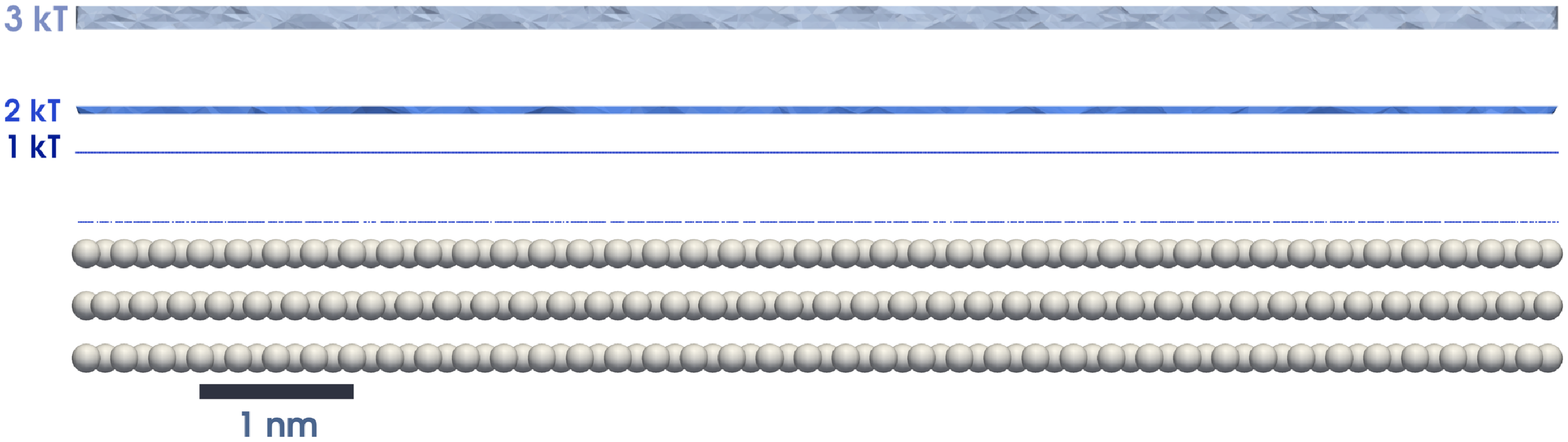}\\
\makebox[0pt]{\raisebox{0.5in}[0pt][0pt]{\raggedleft(b)}}
\includegraphics[bb=150 0 430 842,clip,angle=90,width=0.95\columnwidth]{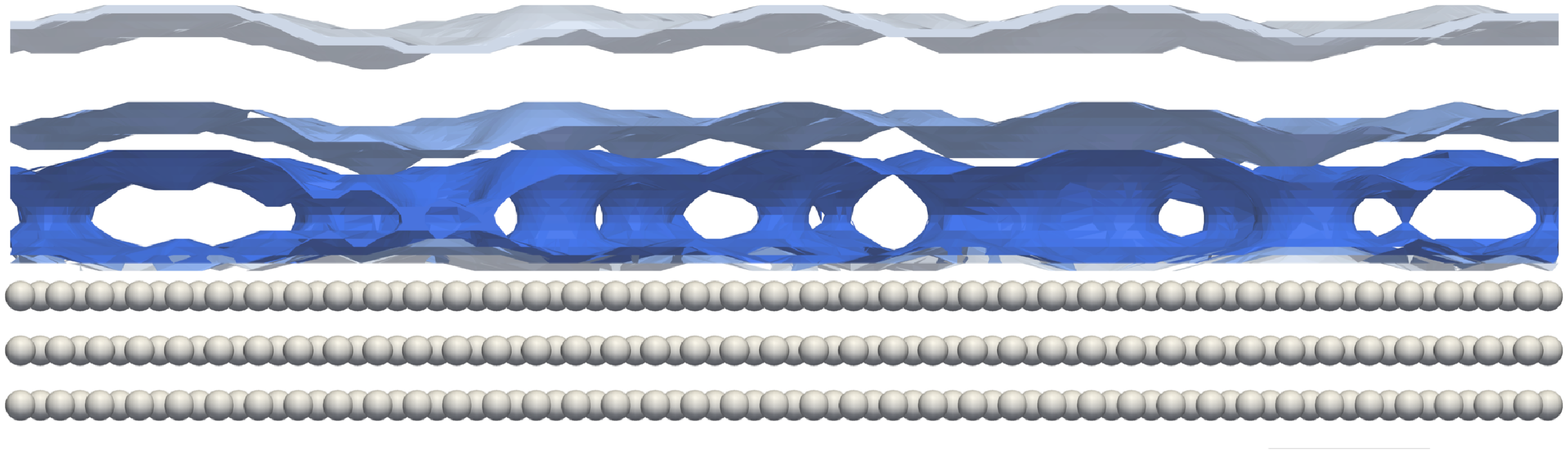}\\
\makebox[0pt]{\raisebox{0.5in}[0pt][0pt]{\raggedleft(c)}}
\includegraphics[bb=150 0 430 842,clip,angle=90,width=0.95\columnwidth]{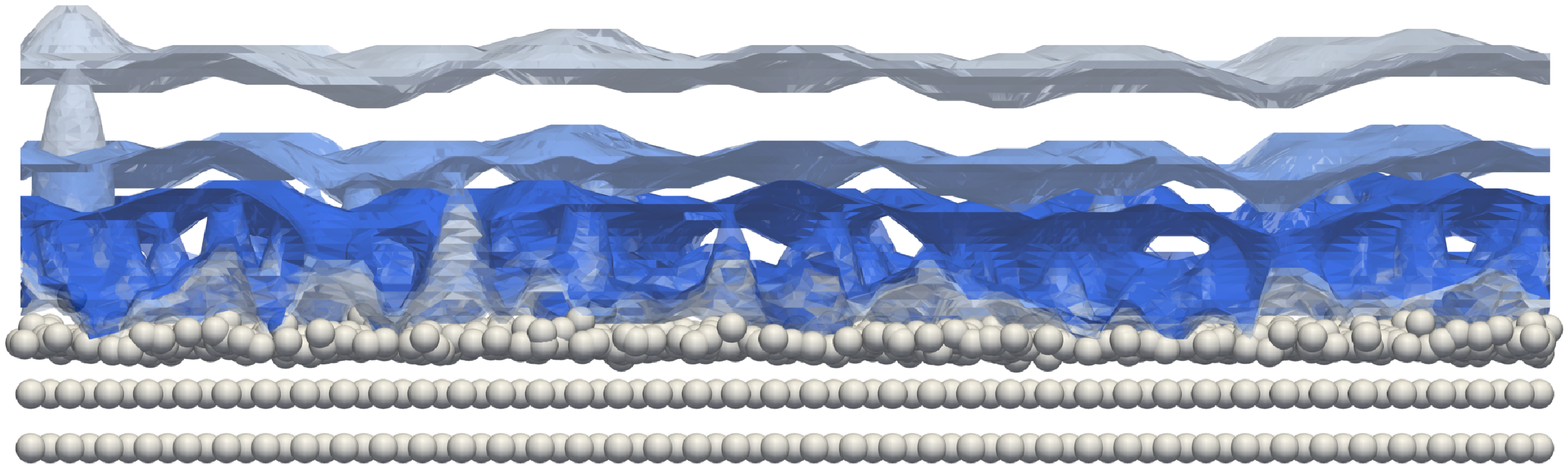}\\
\caption{
Simulation snapshots of a portion of the solid surface
and corresponding potential energy isosurfaces for a single water
molecule at energy $k_BT$, $2 k_BT$ and $3 k_BT$, respectively from
the highest to the lowest color saturation. The vertical position
of the isosurfaces is magnified by a factor 5. Panel (a): standard surface with no atomic displacement; Panel (b):
random quenched functionalization with no atomic displacement; Panel (c): random quenched functionalization with atomic displacement $\xi = \sqrt{k_bT/k}\simeq0.022$ nm.\label{FigScene}}
\end{figure}
\begin{figure}[t]
\includegraphics[width=\columnwidth]{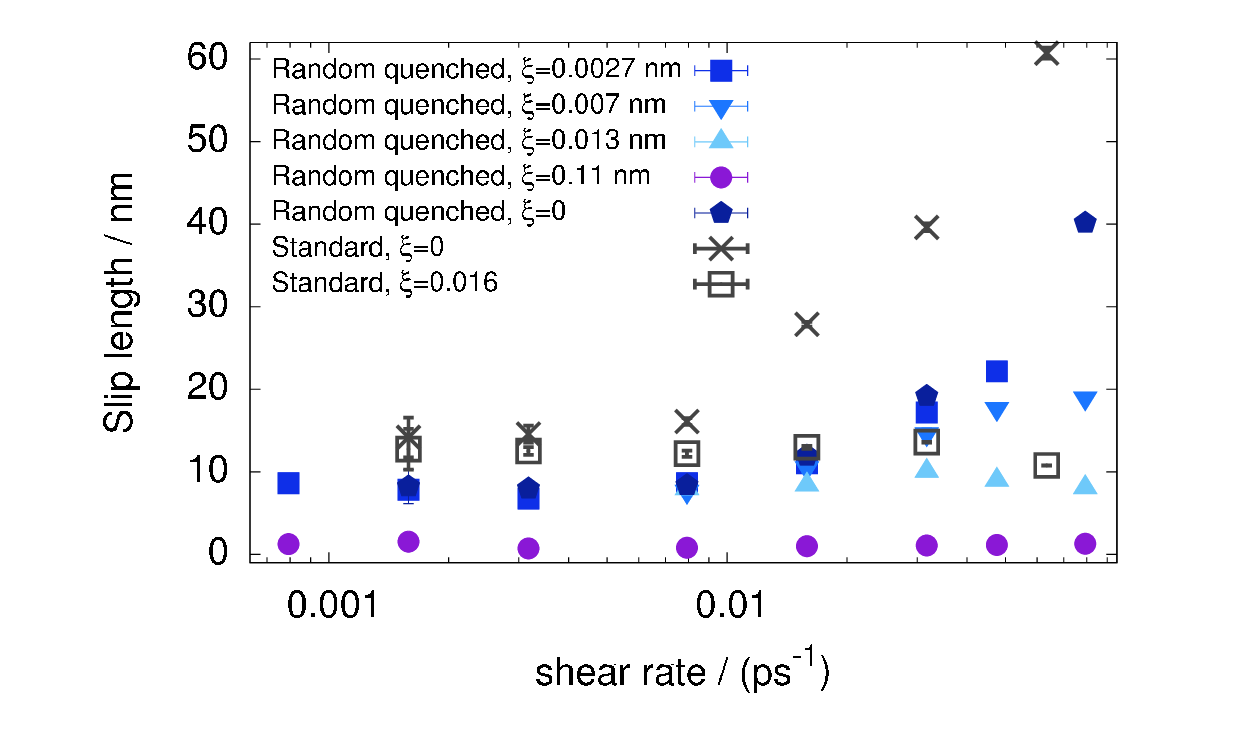} \caption{
Slip length as a function of the shear rate for different surface
types. Very small values
of excursion $\xi$ remove the slip length divergence for both surface
types. Wall atom mass is $m=1.2\times10^3$~amu.}
\label{FigBranches}
\end{figure}
The issue of slip length divergence, however,
is of special interest, as it provides testbed for our
understanding of the fluid/solid interactions, and of the minumum
requirements for the stability of the fluid slip.

In the approximately fifteen years since the publication of work\cite{thompson97},
violations \cite{jabbarzadeh98,priezjev07,martini08}, as well as
confirmations \cite{priezjev04,voronov06,chen06,priezjev06,priezjev11}
of the existence of a slip length divergence have been reported, preventing the emergence of
a clear consensus on the dependence of fluid slippage properties
not only on the imposed shear, but also on the static and dynamic surface
inhomogeneities.  This lack of consensus  calls for further investigations,
since controlling and predicting the behavior of water in nanochannels,
where surface properties have a profound effect on flow dynamics,
stands as a major scientific challenge with many practical
applications in modern science and technology.  Indeed, an accurate
understanding of nanoscale friction phenomena at fluid-solid
interfaces is paramount to the design of micro and nanofluidic
devices aimed at optimizing mass transport against an overwhelming
dissipation barrier \cite{eijkel05,sbragaglia06,benzi06,schoch08,bocquet10}.

The effects of static properties of the fluid-solid interface, such
as the contact angle and surface roughness, on slippage phenomena
have been investigated in depth in the recent
past~\cite{bocquet07,barrat99,barrat99b,huang08,ho11,martini08,thompson97,priezjev11,priezjev06,baumchen10,chinappi10}, not to mention the growing interest in microtextured and superhydrophobic materials\cite{erbil03,lafuma03,choi06,sbragaglia07,bazant08,peters09,feuillebois09,schmieschek12,giacomello12,giacomello12b}.
Recent simulations, for example, suggested that the correlation
between hydrophobicity and high slippage could be less strong than
expected so far, since some model hydrophilic surfaces can show
slip behavior typical of hydrophobic ones\cite{ho11}. Even though mesoscopic
methods have been shown to reproduce quantitatively the results of
the molecular dynamics of simple liquids\cite{horbach06}, the
description at atomistic level of both fluid and surface is crucial
in these kind of investigations, as even very small changes in the
chemical or physical properties of the surface can result in
remarkably different static and dynamic properties of the fluid,
like, for example, in the case of hydroxylated
surfaces\cite{ho11a,qiao12a}.  The role of dynamic properties, such
as the wall flexibility and thermal conductivity, remains however
much less explored~\cite{jabbarzadeh99,priezjev07,hanot13}.  Most
importantly, the geometric effects due to inhomogeneities of flexible
walls, have never been disentangled systematically from the dynamic
ones.  A recent analysis pointed out that all investigations to
date which report absence of a divergent slip length under large
shear were performed with flexible walls capable of exchanging
momentum and energy with fluid~\cite{martini08b}.  From this
observation, and supported by additional numerical calculations and
analytical models, the authors of Ref.~\cite{martini08b} concluded
that heat and momentum transfer from the fluid to the wall are
responsible for the observed saturation of the slip length at high
shear rates, as opposed to the divergent behavior which takes place
using rigid walls.  This picture has been questioned in~\cite{pahlavan11},
where the authors attributed the divergence to the use of a thermostat
applied to the fluid molecules.  Instead of a saturation of the
slip length at high shear rate, these authors report a slippage
drop toward the non-slip condition due to fluid heating.  With the
amount of slippage at high shear rates being reported by different
authors to either vanish, reach a constant finite value, or to
diverge, the present picture appears rather controversial.  More
importantly, no clear consensus has yet emerged on the role of
different physical quantities governing the transition from a finite
to a divergent slip length.

We present here the results of a series of simulations performed
with the aim of discriminating those contributions to the slip
length which are of purely dynamical origin, from those whose origin is rooted in
the surface configurational properties. To this end, we have
performed simulations of water flow in channels with flexible walls 
varying the masses $m$ of the wall atoms and,
independently, the harmonic constant $k$ of the potential that bounds 
wall atoms to their lattice sites (see Fig.~\ref{FigScene}).
According to our findings, this approach proves key to shed new
light on the slip length behavior at high shear rates. More precisely,
we report numerical evidence that it is the presence of an even
extremely tiny amount of disorder in the atomic positions at the
surface, rather than wall flexibility, which proves responsible for
taming slip length divergence and making it shear-independent.

\section{Non-equilibrium Molecular Dynamics simulations of shear
flow} 
\begin{figure}[t]
\includegraphics[width=\columnwidth]{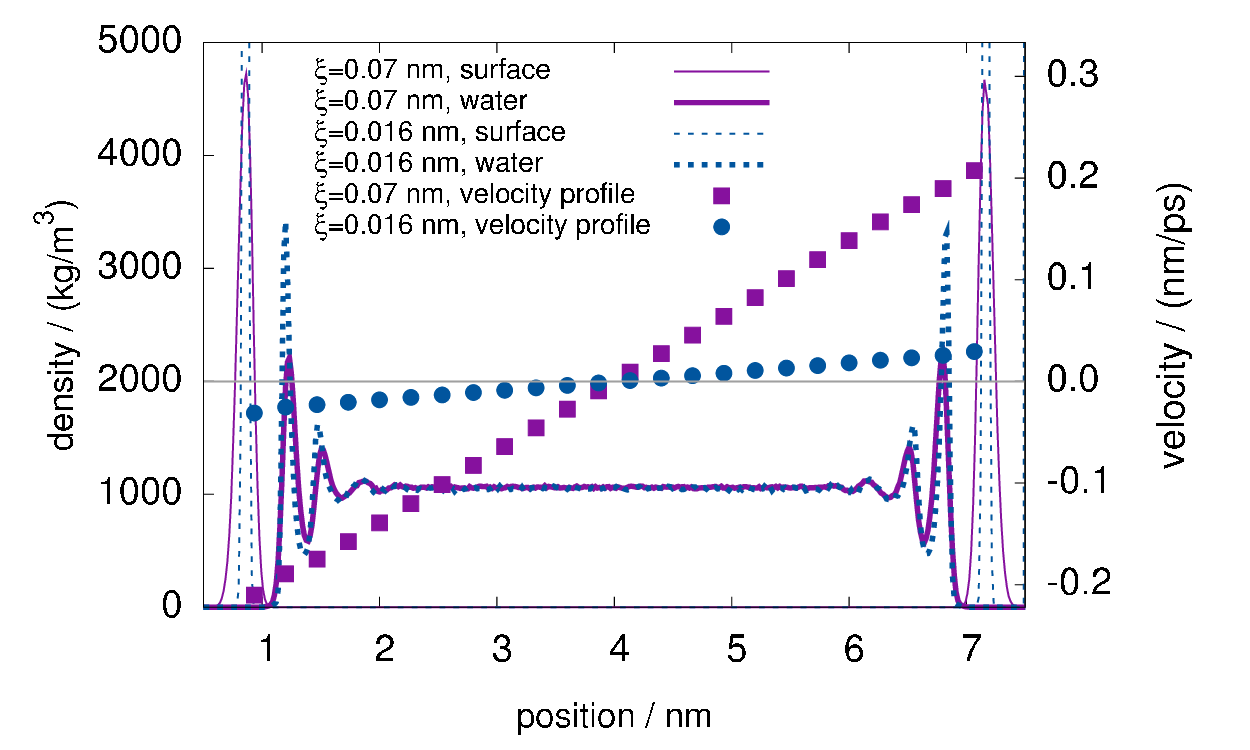}
\caption{Left axis: The density distribution of tethered atoms (thin lines) and water molecules (thick lines) in the random quenched case at $\xi\simeq0.07 $ nm (solid) and $\xi\simeq0.016 $ nm  (dashed). Right axis: water velocity profiles at shear rate $\dot{\gamma}\simeq0.016 \mathrm{ps}^{-1}$ (squares: $\xi\simeq 0.07 $ nm; circles: $\xi\simeq0.016 $ nm).}
\label{fig:profiles} 
\end{figure}

Using a in-house modified version of the Gromacs suite~\cite{hess08} in order to impose Couette flow, 
we simulated a slab of water molecules confined between two
graphite-like walls, each consisting of three atomic layers. The
water molecules are modelled using the SPC/E potential~\cite{berendsen87},
and the electrostatic interaction is calculated using the smooth
particle mesh Ewald method~\cite{essmann95}, together with Yeh-Berkowitz
correction~\cite{yeh99} to remove the contribution from periodic
images along the wall normal.  The distance between the two atomic
planes in direct contact with water defines the channel width
$L=6.306$ nm, while the entire, rectangular simulation cell is
$13.035$ by $16.188$ by $8.0$~nm along $x$, $y$ and $z$, respectively,
with the wall surface normals along the $z$-direction.  The integration
time step for the equations of motion was $\Delta t = 1$~fs, and
the Couette flow was realized by displacing the positions of the
atomic walls every timestep by a fixed amount $\pm u_w\Delta t$,
in addition to the usual movement arising from the integration
of the equation of motion. In this way, a constant walls speed $\pm
u_w$ for the upper and lower wall, respectively, is obtained. At
the stationary state, the velocity profile (see Fig.~\ref{fig:profiles})
is linear over almost the complete extension of the channel, and
deviates from it only at distances from the wall which are smaller
than the size of a water molecule. By fitting the slope of the fluid
velocity, i.e. the effective shear rate,
$\dot{\gamma}_{\mathrm{eff}}=\partial v_x/\partial z$ it is possible
to obtain the slip length from its geometrical definition of the
distance from the wall at which the (extrapolated) fluid velocity
$|\partial v_x/\partial z| (L/2 + \ell_s) $ is equal to the imposed
wall speed $u_w$, \begin{equation} \ell_s = \frac{u_w}{ \left|\partial
v_x/\partial{z} \right|} - L/2.\label{eq:slip_def}\end{equation}
It has to be noted that the effective shear rate $\gamma_{\mathrm{eff}}$
depends implicitly on the channel width $L$ through the fluid/surface
interaction properties, so that Eq.~(\ref{eq:slip_def}), rather
than an expression of the slip length in terms of known quantities,
is merely a conversion between different ways (slip length
and effective shear rate) of measuring the slippage phenomenon: if
the slip length is found, e.g., to be independent on the channel
width, then the effective shear rate must depend on it so
to satisfy Eq.~(\ref{eq:slip_def}), and vice versa.  Note also that
in this work the symbol $\dot{\gamma}$ will be used to denote the
imposed shear rate, $\dot{\gamma}=2u_w/L$, not to be confused with
the effective shear rate $\dot{\gamma}_{\mathrm{eff}}$ that develops
in the fluid.

The viscosity and friction coefficient can be expressed in terms
of the force acting on the surface atoms $f$ as
$\eta=f/\dot{\gamma}_{\mathrm{eff}}$ and  $\lambda=f/u_w
$~\cite{bocquet10}, respectively, but their ratio is independent
from $f$ and equal to $u_w/\dot{\gamma}$, which allows to interpret
the slip length as a measure of the balance between frictional and
viscous forces \begin{equation}\label{eq:slip} \ell_s = \eta/\lambda
- L/2,\end{equation} in analogy with Eq.~(\ref{eq:slip_def}), where
the slip length can be expressed through the ratio between imposed
and effective shear rate. Likewise, the dependence on the channel
width $L$ is not as straightforward as it seems, since the friction
coefficient $\lambda$ depends on the number of water molecules
interacting with the wall and, in turn, on the channel
width.

It could be argued that the appearance of the quantity $L$ in
Eqs.(\ref{eq:slip_def}) and (\ref{eq:slip}) introduces an element
of indetermination in the definition of the slip length if
the walls are rough, as in this case the channel width (that is,
the position of the hydrodynamic boundary) is not well defined.
Partial solutions to this problems have been reported in literature.
For example, in a work that introduces tunable slip length in DPD
simulations~\cite{smiatek08}, and that has been successfully applied
in the simulation of, e.g., nanoconfined electrophoresis~\cite{smiatek10}
and electroosmotic flows~\cite{smiatek09}, it was pointed out that
the slip length and the position of the unknown hydrodynamic boundary
can be determined by performing both a Couette and a Poiseuille
experiment. In the present case, however, the situation is complicated
by the fact that in a Poiseuille flow, in contrast to the Couette
flow, the whole spectrum of shear rates is probed, and this would
have added a possibly even more detrimental indetermination,
since in the present work we focus on the dependence of
the slip length on the shear rate itself. By and large, using
a fixed position for the hydrodynamic boundaries is probably the
best compromise, although in this way the slip length has to be
considered as an effective scale that incorporates a possible shift of
the real hydrodynamic boundary: higher order corrections could be
probably introduced by replacing the geometrical width of the channel
with the effective water slab thickness obtained from the density
profile.  However, these corrections lie outside the scope of this paper.

We investigated two different setups, with respect to the wall-water
interatomic interaction potential.  In the first setup, hereafter
identified as ``standard'', the interaction between wall and the
oxygen atoms in the water molecules is prescribed by a Lennard-Jones
potential $U(r)=C_{12}r^{-12}-C_6 r^{-6} $ with $C_6=2.47512\times
10^{-3}$ (kJ/nm$^6$) and $C_{12}= 2.836328\times 10^{-6}$
(kJ/nm$^{12})$~\cite{vangunsteren96} up to an interatomic distance $r$
of $0.9$ nm. Above that distance, the potential is smoothly switched
to zero at $r=1.2$~nm, using a fourth order polynomial.  The second
setup is a  {\it random quenched} functionalization of the first,
realized by making 40\% of wall atoms purely repulsive ($C_6=0$),
and increasing the interaction strength of the remaining ones by a
factor $\alpha=1.977$.  This generates an intrinsic, mainly {\it
horizontal} inhomogeneity of the surface, e.g. see panel (b) in Fig.~\ref{FigScene}. 
The value of $\alpha$ has been chosen such that water equally wets
the surfaces of the two setups, resulting in comparable macroscopic
contact angles~\cite{werder03}. Given the known dependence of
slip length on contact angle~\cite{huang08}, equal wetting is
prerequisite to comparing the slip length along two microscopically
different surfaces and understanding  to what extent the slip
length is affected by different types of surface inhomogeneities,
rather than by the contact angle.

For each of the surface type just described, we simulated a flexible
and a rigid variant. The flexible variant is  realized by tethering
the atoms of each wall's outermost layer to their respective lattice
sites via a harmonic potential $U_k(r)=\frac{1}{2} k r^2$. The
spring constant $k$ defines the characteristic excursion
$\xi=\sqrt{k_BT/k}$  of the tethered atoms, so that  a small value
of $k$ leads to a high effective inhomogeneity. The density profiles
of water molecules and surface atoms across the channel for two
selected values of $\xi$ are shown in Fig.\ref{fig:profiles}.
Both profiles are more sharply peaked in the small excursion case, but
the effect is significant only next to the surface, and diminishes
greatly by the second density peak.  In both cases the flow velocity
is well represented by the Couette flow solution, even in regions
near the wall where the density of water becomes negligible.

The rigid variant, on
the contrary, is realized by freezing all surface atoms either at
their lattice site position (corresponding effectively to $\xi=0$), or in a
configuration taken from the equilibrium trajectory of the flexible
variant with finite $\xi$, which we will denote by $m=\infty$.  Both
mobile and fixed wall atoms interact with water only, and not among
themselves. By varying the mass $m$ of the tethered molecules, it
is possible to achieve a separate control over dynamic properties,
as the value of $m$ does not influence configurational properties
(e.g.: particles distribution, free energies, contact angles), but
only dynamical ones, like water-surface friction and thermal
conductivity.  As a result, we are able to assess the dependence
of the slip length on dynamic, static and non-equilibrium properties,
controlled by the mass and mean excursion of the wall atoms and by
the imposed shear.
While this model is surely not a completely realistic representation
of a real surface, it allows us to explore in general terms the dependence
of slip length as a function of the magnitude of surface inhomogeneities.
It is worth noticing that in this work we investigate fluctuations
with an extension that is not large enough to model what usually 
goes under the name of surface roughness. Instead, we are reaching
with continuity the limit of very small surface fluctuations, such
as those caused by thermal motion.

Fig.~\ref{FigScene} shows the potential energy isosurfaces for three
different cases. While the spatial functionalization (b) already
introduces a noticeable perturbation of the energy landscape, the addition
of an even tiny spatial displacement of the surface
atoms (c) generates the largest perturbation. 

\section{The slip length divergence and its regularization}
As a first step, with the mass of surface atoms held at a constant
$m=1.2\times10^3$~amu (to efficiently integrate the equations of motion of
the tethered atoms even with the largest spring constant) we
investigated the shear rate dependence of the slip length for
different displacements $\xi$. Providing useful information about the scales which come into play in the determination of slippage properties, we present 
an overview of this dependence for both the standard and randomized surfaces in Fig.~\ref{FigBranches}.
Five different cases were explored for the random quenched surface,
between the values $\xi=0$ and $\xi\simeq 0.11$ nm ($k=200$ kJ/mol
nm$^{-2}$), the former mimicking an infinite spring constant,  and
the latter corresponding to the one proposed as optimal
in~\cite{martini08b}. The smallest but still finite value,
$\xi\simeq0.0027$ nm, was chosen to model a nearly flat wall which
yet retains dynamical features.

\begin{figure}
\includegraphics[width=\columnwidth]{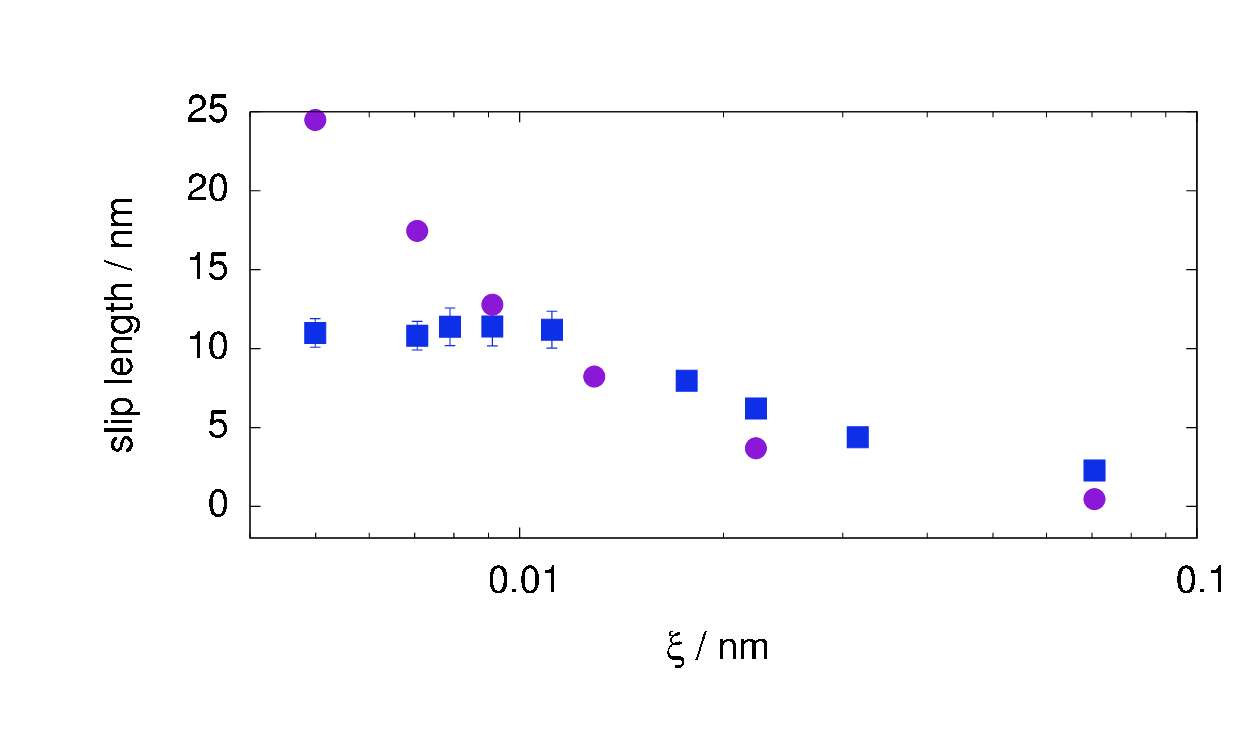}
\caption{Slip length as a function of the excursion $\xi=\sqrt{k_BT/k}$ for two different values of shear rate (squares: $\dot{\gamma}=0.016$ ps$^{-1}$; circles: $\dot{\gamma}=0.08$ ps$^{-1}$). Tethered atoms have a mass of $1.2\times10^3$ amu. \label{FigTransition}}
\end{figure}

When the wall atoms are perfectly flat ($\xi=0$), both surface types
show a tendency towards a larger --- in fact {\it divergent} ---
slip length at increasing shear. It takes only sub-nanoscopic
excursions of wall atoms to remove the divergence and make the slip
length shear independent. The slip length is shear-independent
(constant) also for large values of $\xi$, however the constant
value does increase with the spring constant $k$ (decreasing excursion
$\xi$) to a limiting value of about $8$ nm for the random quenched
surface and a slightly larger one for the standard surface. The
maximum limiting value for shear-independent slip lengths is reached
at $\xi_c \simeq 0.01$~nm. For  smaller mean excursions, the slip
length is no longer shear-independent and a divergent-like behavior
is observed instead. At first glance, a characteristic excursion
$\xi=0.01$~nm might appear to be surprisingly small, but we note
that several other important scales in this system, such as the
width of the $1 k_BT$ energy basin of a water molecule in the
direction normal to the surface (see Fig.~\ref{FigScene}), and the
distance $\delta\simeq0.03$~nm travelled by a water molecule during
its translational relaxation time of $\tau_r\simeq0.1$~ps~\cite{toukan85}.
These, however, are only considerations about the orders of magnitude
of several process occurring in the system, and a true microscopic
explanation of the reason why a length of 0.01 nm is sufficient to trigger the
transition from a divergent to a constant slip length is missing.

The transition from shear-independent to divergent slip length is
made even more evident by plotting the slip length as a function
of $\xi$ (Fig.~\ref{FigTransition}) for two selected shear rates,
$\dot{\gamma}\simeq0.016$ ps$^{-1}$ at the upper end of the low
shear-rates plateau, and $\dot{\gamma}\simeq0.08$ ps$^{-1}$ well
into the shear region where divergence is observed.  At the lower
shear rate, saturation is reached for excursions smaller than
$\xi_c\simeq 0.01$ nm, whereas the high-shear rate slip length keeps
increasing, seemingly unaffected.  

From the analysis presented thus far, the role of wall flexibility
remains unclear, as this dynamical aspect was not separated out
from the other features.  For this reason, we selected two extreme
values of excursion ($\xi=0.0027$ and $0.11$~nm) and calculated the
slip length for several values of the tethered masses $m$, including
the case of atoms fixed in one equilibrium configuration, $m=\infty$,
as described before.

\begin{figure}[t] 
\includegraphics[width=\columnwidth]{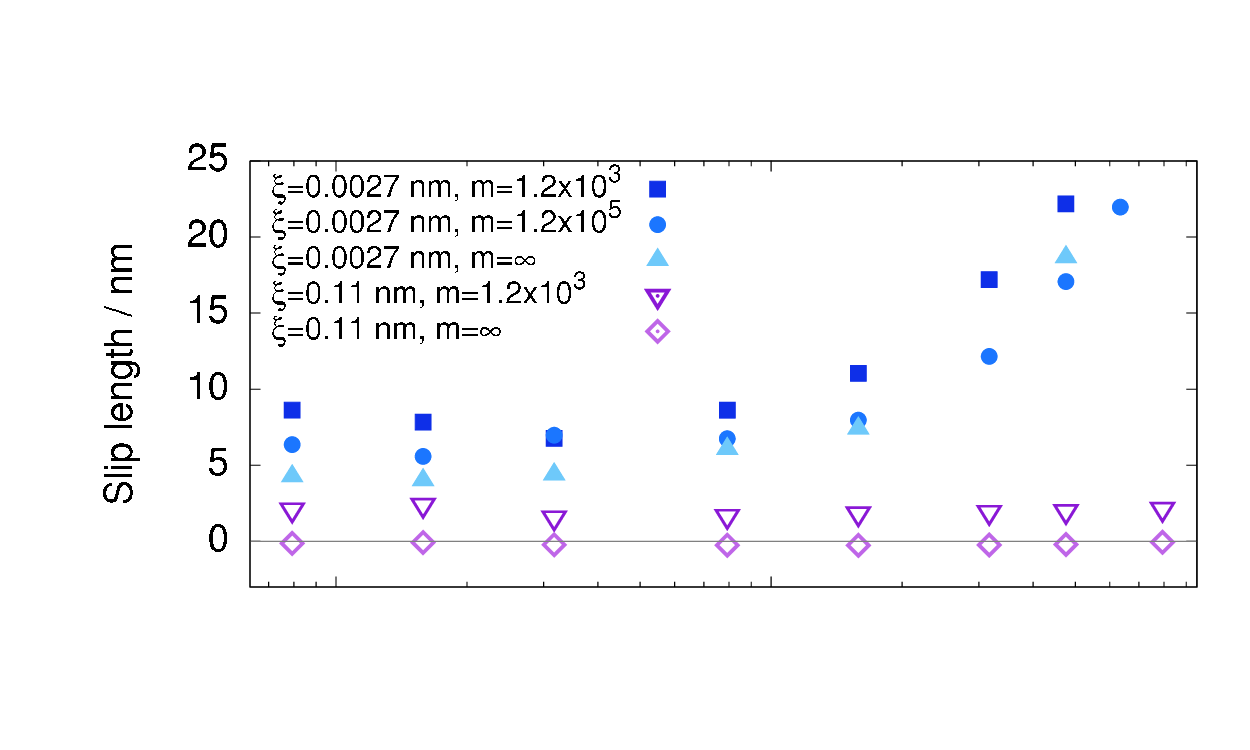}
\includegraphics[width=\columnwidth]{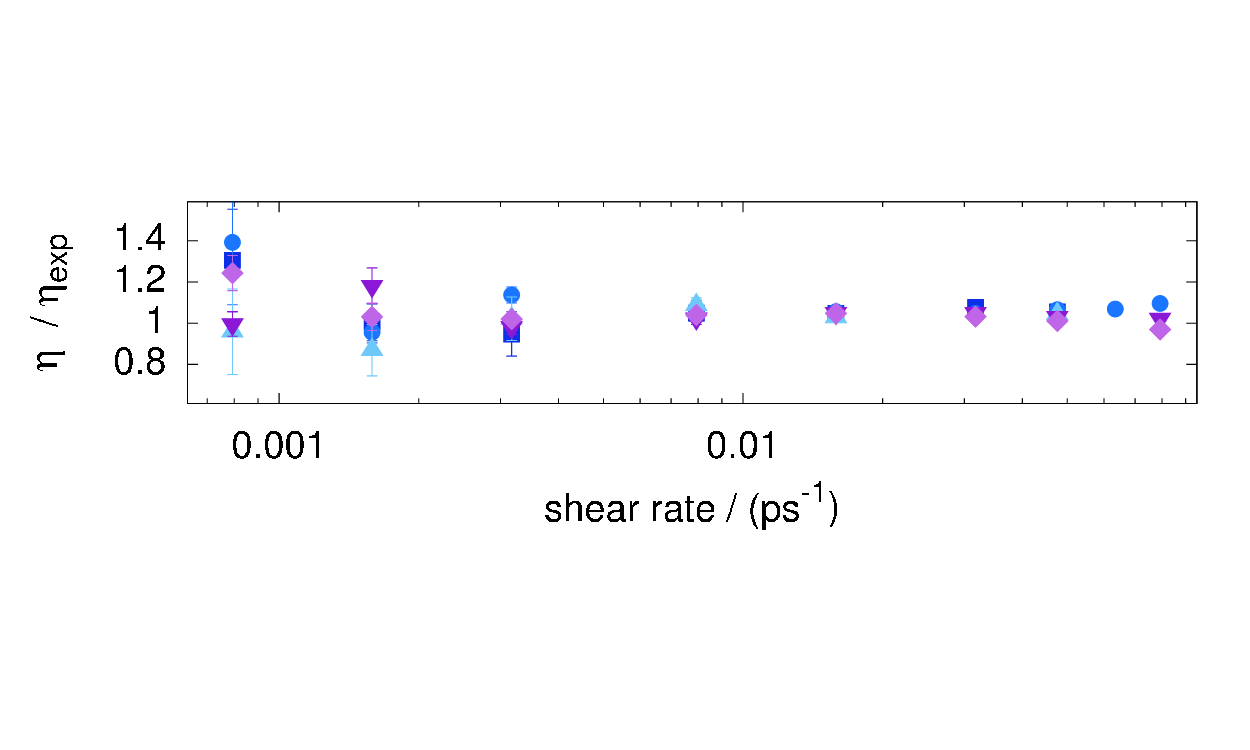} 
\caption{Slip length as a function of the shear rate in the random quenched case, for $\xi\simeq 0.0027$ nm (squares: $m=1.2\times10^3$ uma; circles: $m=1.2\times10^5$uma; upper triangles: $m=\infty$) and for $\xi\simeq 0.11$ nm (lower triangles: $m=1.2\times10^3$uma; rhombs: $m=\infty$). Lower panel: water viscosity as a function of shear rate. The experimental value $\eta_\mathrm{exp}=515.5$ kJ/mol/ps/nm$^3$ at 300K has been interpolated from reference data\cite{kestin78a}.\label{FigMaster}} \end{figure}

Results from these calculations are shown in Fig.~\ref{FigMaster}.
At low shear rates ($\dot{\gamma}<10^{-2}$ ps$^{-1}$)  both values of $\xi$ 
demonstrate marked changes to the slip length with the mass $m$. 
The small excursion case $\xi=0.0027$~nm halves in slip
from the lowest mass $m=1.2\times10^3$~amu to $m=\infty$, and in the $\xi=0.11$~nm case the slip length is reduced to zero at $m=\infty$.
The transition from thermally insulating walls ($m=\infty$) to
conducting walls is therefore associated to a substantial increase
in slip length. In other words a flexible wall reduces friction
on the fluid (therefore, increasing the slip length) by being capable
of adjusting locally to the flux of water molecules. In the extreme
case of $m=\infty$, the slip length divergence is indeed removed,
provided that the excursion $\xi$ is larger than $\xi_c$. This
provides evidence that the flexibility can not be responsible for
the regularization of the slip length at high shear-rates, and acts
instead in the opposite direction, by increasing, instead of
decreasing, the slip length.

It is important to note that these are entirely surface-related
effects. The viscosity of water $\eta$ is the only bulk property
that appears in the definition of the slip length for a Couette
flow, Eq.~(\ref{eq:slip}), and it does not show any dependence on
shear, or spring constant, as shown in Fig.~\ref{FigMaster}, bottom
panel. Therefore,  we argue that the transition from a divergent
to a shear-independent slip length is determined not by the wall
flexibility, but rather by the configurational disorder induced by
tiny displacements of the wall atoms in the direction normal to the flow. 

 For Lennard-Jones liquids,
Priezjev already noticed~\cite{priezjev07} that very small, static
displacements of the atomic surface atoms (already 7\% of the
Lennard-Jones diameter) can have a dramatic influence on the slip
length using thermal, random or periodic displacements. The systematic
analysis performed here puts previous results in perspective,
regarding the important case of water slippage. In particular, to
get an appreciation for the physical realism of the characteristic
excursion lengths discussed here, we compare them to displacements
obtained from the effective spring constants of real graphite at
300~K measured by inelastic X-ray scattering \cite{mohr07}, which
correspond to about $0.006$~nm for nearest neighbors, and are much
larger for second neighbors. Thermal fluctuations alone are therefore
expected to be sufficient to attain a shear-independent slip length
for water.

\section{Conclusions}
We have performed non-equilibrium Molecular Dynamics simulations
of water flow in nano-channels with separate control of the flexibility
and static inhomogeneity of the wall. The simulations show that the
disappearance of the divergence of the slip length at high shear
rates, formerly ascribed to wall flexibility, is due instead to the
excursion of wall atoms from the ideal horizontal plane.  Moreover,
these results show that atomic displacements as small as those due
to thermal motion at room temperature are sufficient to regulate
the divergence of the slip length. We conclude that, even
in the absence of surface imperfections such as point defects or
dislocations, high-shear water flows in nanoscale channels should
not exhibit any divergent slip length; that thermal motion of the
wall atoms is sufficient to tame such divergence.

We acknowledge M. Chinappi \&  D. Lohse for useful discussions and
E. M. Foard for a careful reading of the manuscript.  M. Sega, M.
Sbragaglia \& L. Biferale acknowledge ERC-DROEMU c.n. 279004 for
support.  M. Sega acknowledges FP7 IEF p.n. 331932 SIDIS for support.

\footnotesize{

\providecommand*{\mcitethebibliography}{\thebibliography}
\csname @ifundefined\endcsname{endmcitethebibliography}
{\let\endmcitethebibliography\endthebibliography}{}

}

\end{document}